\documentclass[aps, prd, twocolumn, showpacs, superscriptaddress, groupedaddress]{revtex4} 
\usepackage{graphicx}	
\usepackage{amssymb}
\usepackage{dcolumn}
\usepackage{color}
\usepackage{amsmath}
\usepackage{subfigure, rotating, bm, array}
\usepackage[pagebackref=false, colorlinks=true]{hyperref}
\hypersetup{
linkcolor=blue,     
citecolor=blue,     
urlcolor=blue}      
\begin{document}
\title{Study of Orbital Dynamics in Singular and Regular Naked Singularity Space-times}
\author{Avisikta Ghosh}
\email{avisikta.1ghosh@gmail.com}
\affiliation{PDPIAS, Charusat University, Anand, GUJ 388421, India}

\date{\today}

\begin{abstract}
The universe is filled with various compact objects and the most attractive of them are the black holes and singularity. But it is also known that at the singularity density becomes so infinitely high that the present physics knowledge breaks down. Thus, the singularity remains a flaw of the present theories.
Several methods have been exercised to resolve the singularity. One such mathematical method is through conformal transformations. This paper deals with regularizing a naked singularity space-time using conformal transformations, further studying and comparing its time-like orbits with that of the naked singularity space-time.


\end{abstract}

\maketitle

\section{Introduction}

During the gravitational collapse of a massive matter cloud, depending upon the initial conditions the final state of the collapse becomes either a black hole or a naked singularity. In either of the case, the final state always yields a singularity. But at the singularity, the present physical theories vanish, therefore, serving it as a limitation of the general theory of relativity. \\
The observations of the stellar motions (i.e., motion of S2, S55 stars etc) by the GRAVITY collaboration and the UCLA Galactic Center Group have sparked the interest to investigate the central regions of the galaxies. Recently, many significant researches on studying the time-like orbits around black holes and naked singularity space-times have been done \cite{padp}, \cite{pddapv}. Due to the presence of an event horizon around the singularity in a black hole, the information from the singularity cannot reach the outside observer. But as a naked singularity lacks an event horizon, it is possible for the information from the singularity to reach the outside observer. Thus, studying the effects of regularization on a naked singularity space-time can help us to gain significant results.\\ 
One way to resolve the singularity is by applying conformal transformations to the singular space-time and then studying and comparing the time-like orbits and precession of the regularized space-time with that of the singular space-time.\\
This paper is organised as follows. In section II, we will discuss about regularization of a space-time using conformal transformation. Then we will further discuss about regularizing a Schwarzschild black hole which is also discussed in \cite{bambi} and then study and plot its time-like orbits. Further, we will discuss about regularizing a naked singularity space-time \cite{adpp} by using a suitable conformal factor and calculate its corresponding Ricci and Kretschmann scalars. In section III, we examine the effective potential and time-like orbit equation for the naked singularity and regularized naked singularity space-times. The section IV discusses about the precession and shape of the bound time-like orbits and their corresponding orbit plots for the naked singularity and regularized naked singularity space-times. In section V, we conclude with discussing the important results of this paper.

\section{Regularization of a Space-time}
Conformal transformations serve as an appealing proposal to resolve the space-time singularities. The conformal transformation of any space-time is given by,
\begin{equation} \label{S}
    d\tilde{S}^2 = S(r)\hspace{0.1cm} dS^2,
\end{equation}
where S(r) is the conformal factor which is a function of the space-time point. The causal structure remains intact after conformal transformation of a space-time \cite{john}. Also, there can be more than one conformal factor to regularize a particular space-time \cite{bambi}. Thus, by choosing a conformal factor S(r) for the space-time metric, this method can solve the problem of space-time singularity \cite{bambi} \cite{zhang}.

\subsection{Regularizing Schwarzschild Black Hole}
Line element of Schwarzschild black hole space-time is given by,
\begin{equation}
\begin{split}
    dS^{2}_{SCH} = -\Big(1 - \frac{2M}{r}\Big) dt^2 \hspace{0.05cm}+\hspace{0.05cm} \Big(1 - \frac{2M}{r}\Big)^{-1} dr^2 \hspace{0.05cm}+\hspace{0.05cm} r^2 d\theta^2 \\ \hspace{0.05cm}+\hspace{0.05cm} r^2 sin^2\theta d\phi^2
\end{split}
\end{equation}

From \cite{bambi}, a regular Schwarzschild black hole space-time can be given by,

\begin{equation}\label{RegSCH}
d\tilde{S}^{2} = \Big(1 + \frac{L^4}{r^4}\Big) \hspace{0.1cm} dS^{2}_{SCH},
\end{equation}

where $\Big(1 + \frac{L^4}{r^4}\Big)$ is the conformal factor. The Ricci scalar and the Kretschmann scalar for the regularized Schwarzschild black hole are \cite{bambi},

\begin{equation}
    R = -\frac{12 L^4 r (L^4(r - 4 M) + r^4(3 r - 8 M))}{(L^4 + r^4)^3}
\end{equation} 

\begin{equation}
\begin{split}
K = \frac{16 r^2}{(L^4 + r^4)^6} \times [L^{16}(39M^2 - 20Mr + 3r^2) \\+ 2L^{12}r^4(66M^2 - 32Mr + 3r^2) +\\ L^8r^8(342M^2 - 284Mr + 63r^2) + 12L^4M^2r^{12} + 3M^2r^{16}]
\end{split}
\end{equation}
We can observe that both the scalars have become zero or finite quantities at the center r=0, thus, regularizing the Schwarzschild black hole.\\
In this paper we are interested in regularizing a naked singularity space-time. Then we will study its orbital dynamics and compare the results before and after regularizing the naked singularity space-time by conformal transformation.
\subsection{Regularizing naked singularity space-time}
We are going to regularize the naked singularity space-time given in \cite{adpp}, using conformal transformation. To do so, we need to find a suitable conformal factor which on applying the conformal transformation makes the scalars zero or finite.\newline 
The naked singularity space-time \cite{adpp} is given as
\begin{equation} \label{naked}
\begin{split}
d{S}^2 =  - \frac{dt^2}{\Big(1+\frac{M}{r}\Big)^2} +  \Big(1+\frac{M}{r}\Big)^2 dr^2 \\ + r^2 d\Omega^2 
\end{split}
\end{equation}
The conformal factor for eq. (\ref{naked}) is,
\begin{equation} \label{confactor}
    S(r) = \Big(1+\frac{L^2}{r^2}\Big)
\end{equation}
The regularized naked singularity space-time can be written as eq. (\ref{S}). Thus, we get,
\begin{equation} \label{regular}
d\tilde{S}^{2} = \Big(1+\frac{L^2}{r^2}\Big)  \hspace{0.1cm} dS^{2},
\end{equation}
where L is a parameter with dimensions of length and is of the order of the Planck scale. At asymptotic limit, the conformal factor becomes unity.\\
Now, the corresponding Ricci scalar and Kretschmann scalar are,
\begin{equation}
\begin{split}
    R = \frac{1}{2\Big(1+\frac{L^2}{r^2}\Big)^3r^2(M+r)^4} \times\\ \Big[12L^4+12L^2M^2\Big(1+\frac{L^2}{r^2}\Big)\\+4M^4\Big(1+\frac{L^2}{r^2}\Big)^2+\frac{12L^4M^2}{r^2}+\\\frac{24L^4M}{r}+16M^3\Big(1+\frac{L^2}{r^2}\Big)^2r-\\12L^2\Big(1+\frac{L^2}{r^2}\Big)r^2\Big]
\end{split}
\end{equation}
and
\begin{equation}
    K = \frac{4}{(M+r)^8 (L^2+r^2)^6} \times (Polynomial \hspace{0.1cm} of \hspace{0.1cm} L, M, r)
\end{equation}
The values of R and K at the center $r = 0$ are,
\begin{equation}
    \lim_{r\to0} R = \frac{2}{L^2}
\end{equation}
\begin{equation}
and \hspace{0.1cm} \lim_{r\to0} K = \frac{4}{L^4}
\end{equation}
From the above expressions we can see that both the scalars provide a finite value at r = 0 and thus, the space-time is regularized.

\section{Time-like Orbit Equations in Singular and Regular Space-times}
\subsection{Time-like Orbits of Regular Schwarzschild Black Hole}
For the regularized space-time in eq. (\ref{RegSCH}), considering the equatorial region and using the conserved $\gamma$ and l we get,
\begin{equation}
    \begin{split}
        u^t = e \Big(1 + \frac{L^4}{r^4}\Big)^{-1} \Big(1 - \frac{2M}{r}\Big)^{-1}\\
        and \hspace{0.3cm} u^\phi = \frac{l}{r^2} \Big(1 + \frac{L^4}{r^4}\Big)^{-1}
    \end{split}
\end{equation}
The effective potential is,
\begin{equation}
    V_{eff} = \frac{1}{2} \bigg[\frac{l^2}{r^2} \Big(1 - \frac{2M}{r}\Big) + \bigg(1 + \frac{L^4}{r^2}\Big) \Big(1 - \frac{2M}{r}\Big) - 1\bigg]
\end{equation}

\begin{figure*} 
\centering
\subfigure[] {\includegraphics[scale=0.5]{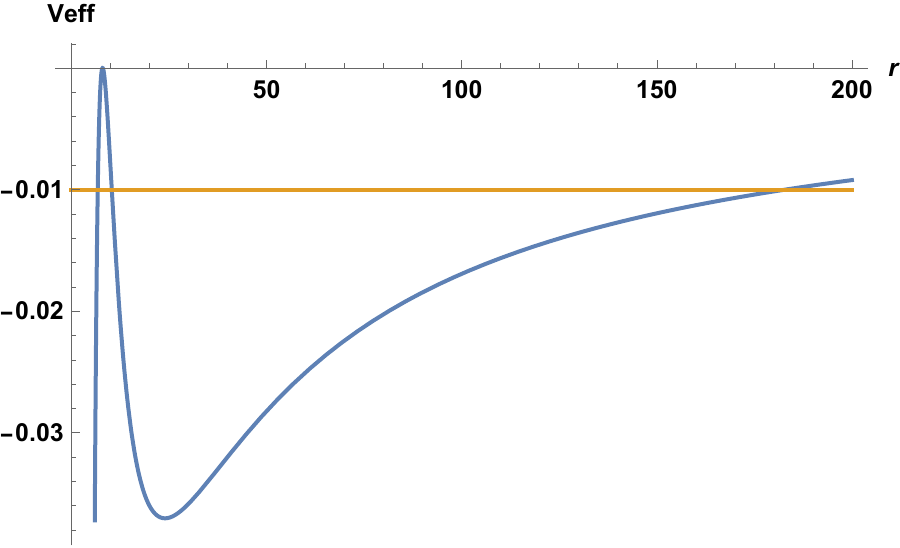}\label{VeffSCH}}
\hspace{0.2cm}
\subfigure[]{\includegraphics[scale=0.55]{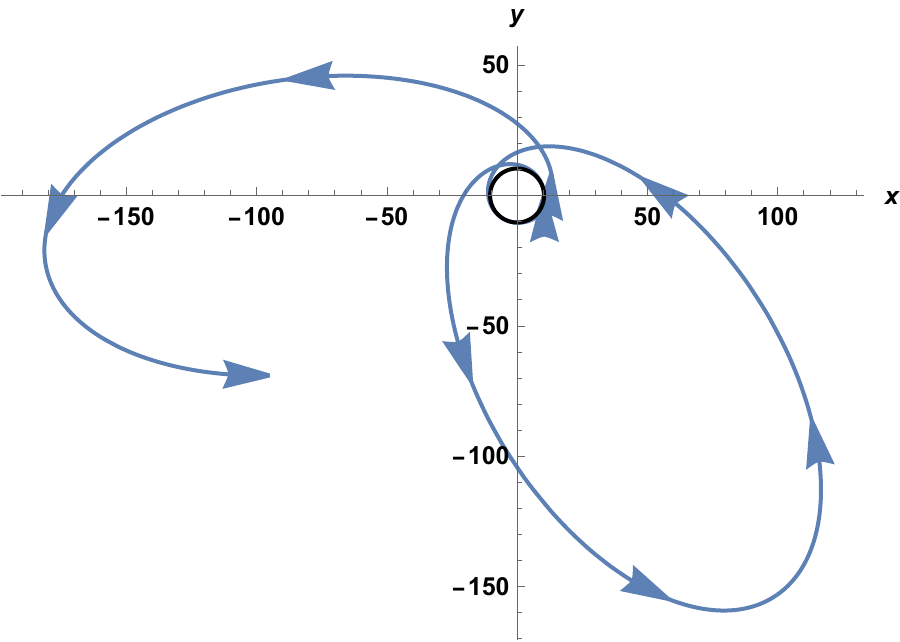}\label{OrbitSCH}}
\caption{The figures in (a) and (b) show the effective potential plot and the time-like orbits plot for regular Schwarzschild black hole}
\end{figure*}

The plot of effective potential for the regularized Schwarzschild space-time is given in fig. (\ref{VeffSCH}). 
The time-like orbit equation for the regular Schwarzschild space-time in eq. (\ref{RegSCH}) is,
\begin{equation}\label{orbiteqSCH}
    \begin{split}
        \frac{d^2 u}{d\phi^2} + u(1-2Mu) - Mu^2 + \frac{2L^4u^3}{l^2}(1-2Mu) - \\
        \frac{M}{l^2}(1+L^4u^4) = 0
    \end{split}
\end{equation}
The orbit plot for eq. (\ref{orbiteqSCH}) is given in fig. (\ref{OrbitSCH}).

\subsection{For Naked Singularity Space-time}
\begin{figure*} 
\centering
\subfigure[] {\includegraphics[scale=0.54]{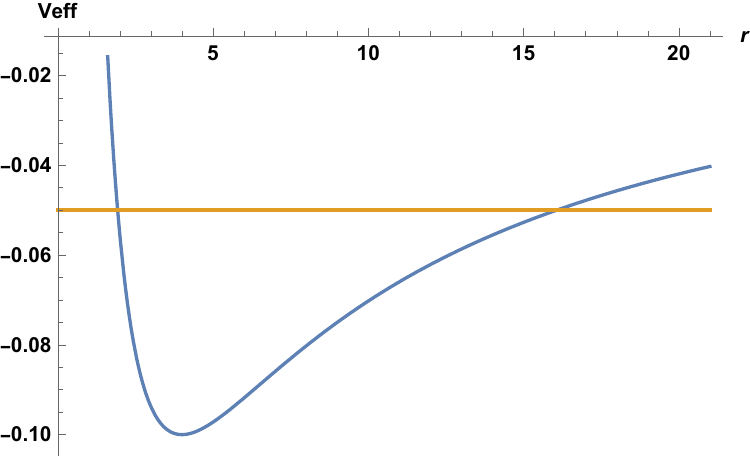}\label{VeffWCF}}
\hspace{0.5cm}
\subfigure[]{\includegraphics[scale=0.55]{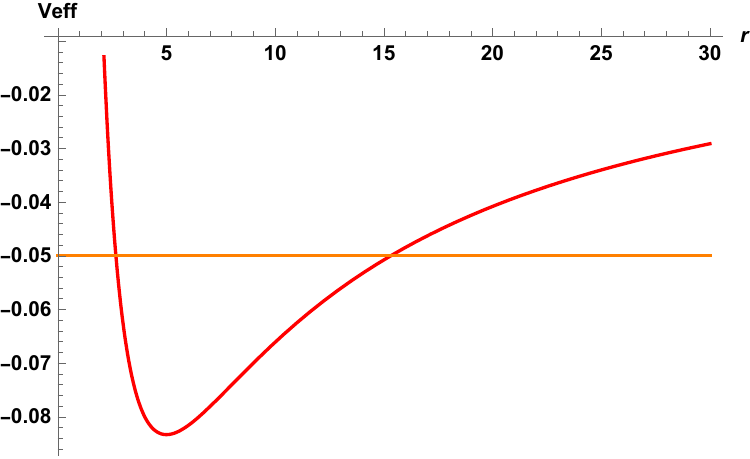}\label{VeffCF}}
\hspace{0.2cm}
\subfigure[]{\includegraphics[scale=0.55]{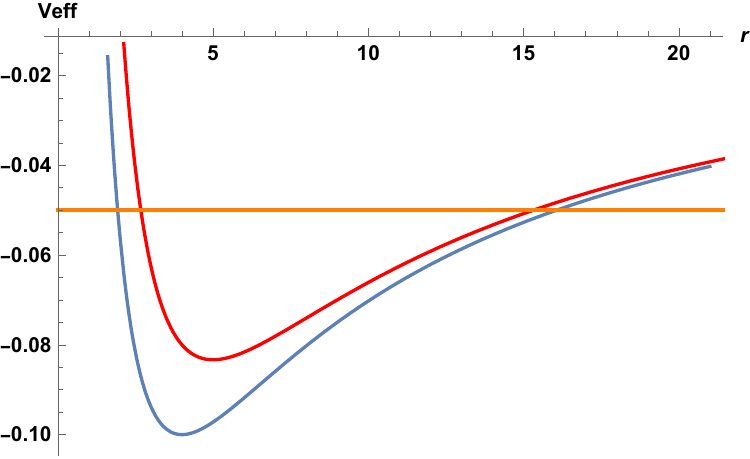}}
\caption{The figures in (a) and (b) show the effective potential plots of singular and regular naked singularity space-times respectively. Figure (c) shows the plot comparison between the effective potential plots in (a) and (b).}
\end{figure*}

\begin{figure*}
\centering
\subfigure[Particle orbit in naked singularity space-time with $l=1.5$ and $M=1$]
{\includegraphics[scale=0.5]{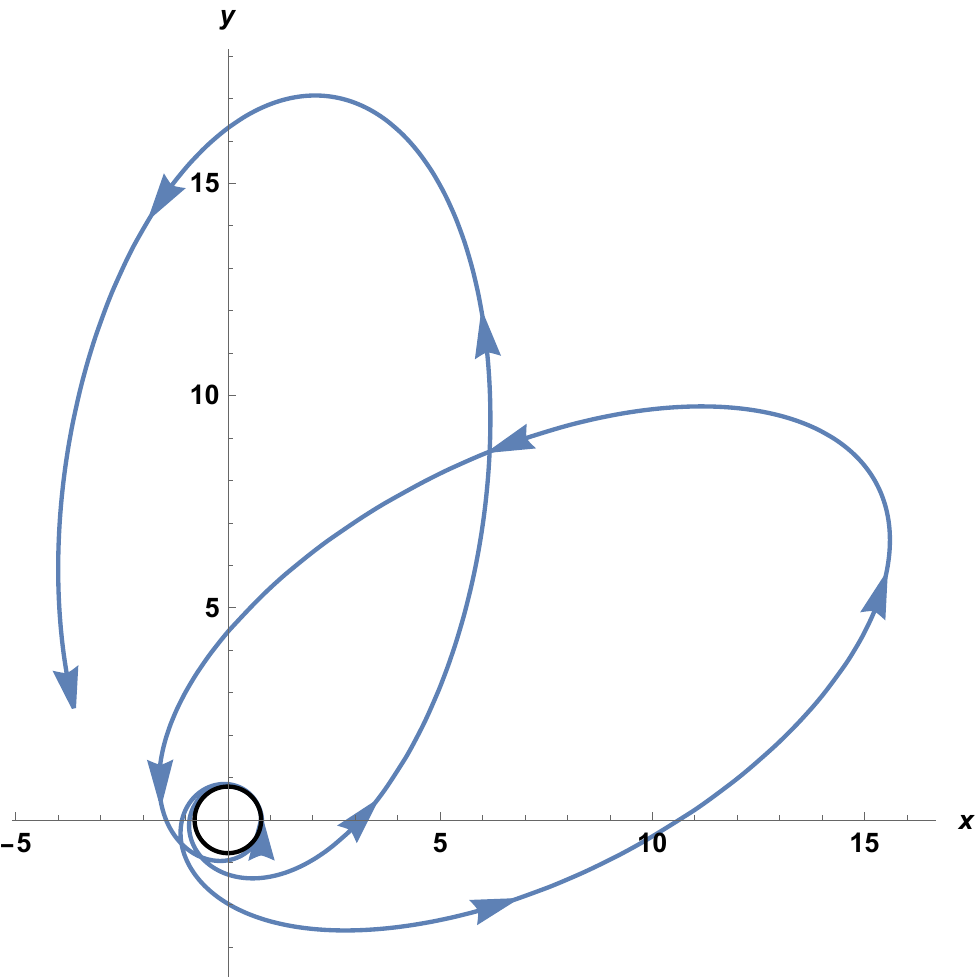}\label{WCFPlot}}
\hspace{0.2cm}
\subfigure[Particle orbit in regular naked singularity space-time with $l=1.5$, $M=1$ and $L=0.6$]
{\includegraphics[scale=0.5]{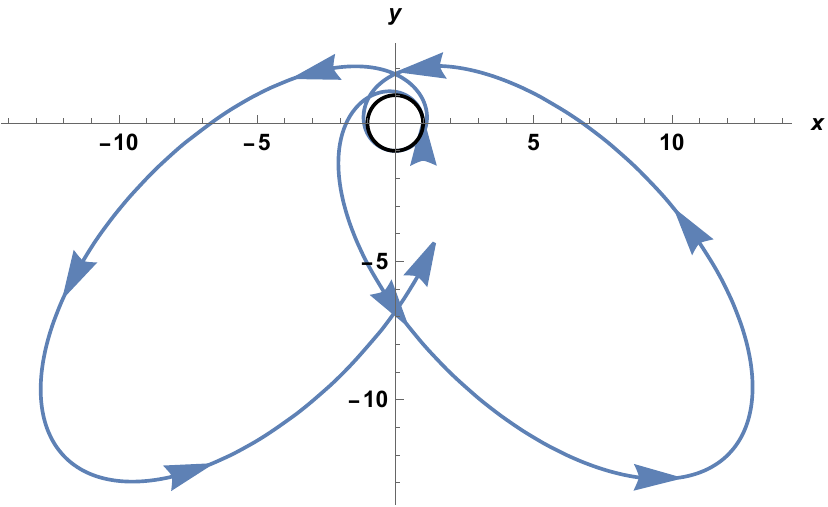}\label{CFPlot}}
\caption{Above figures show particle orbits in singular and regular space-times given in eq.(\ref{naked}) and eq.(\ref{regular}) respectively.}
\end{figure*}

As the naked singularity space-time in \cite{adpp} is static and spherically symmetric, the energy $(\gamma)$ and angular momentum (l) per unit rest mass of a freely falling particle are conserved. The general form of these conserved quantities using the killing vectors can be written as, 
\begin{equation}
\begin{split}
     u^t \hspace{0.1cm} \zeta^t_{(t)} \hspace{0.1cm} g_{tt} = \gamma \\
     u^\phi \hspace{0.1cm} \zeta^\phi_{(\phi)} \hspace{0.1cm} g_{\phi\phi} = l
\end{split}
\end{equation}
Here, we have considered equatorial region, so, $\theta = \frac{\pi}{2}$.
So for the naked singularity space-time from \cite{adpp},
\begin{equation}
dS^2 = -\frac{dt^2}{\Big(1+\frac{M}{r}\Big)^{2}} + \Big(1+\frac{M}{r}\Big)^2 dr^2 + r^2 d\Omega^2,
\end{equation}
we get,
\begin{equation}
    \begin{split}
        u^t = \frac{\gamma}{\Big(1+\frac{M}{r}\Big)^{-2}}\\
    and \hspace{0.3cm} u^\phi = \frac{l}{r^2}
    \end{split}
\end{equation}
As a particle always follows time-like geodesics, $u^\alpha u_\alpha=-1$. The total energy can be written as,
\begin{equation}
    E = \frac{1}{2}\Big[\Big(\frac{dr}{d\tau}\Big)^2 + \frac{l^2}{r^2}\Big(1+\frac{M}{r}\Big)^{-2}+\Big(1+\frac{M}{r}\Big)^{-2}-1\Big]
\end{equation}
Thus, the effective potential obtained for this space-time is,
\begin{equation}
   V_{eff} = \frac{1}{2}\Big[\frac{l^2}{r^2}\Big(1+\frac{M}{r}\Big)^{-2}+\Big(1+\frac{M}{r}\Big)^{-2}-1\Big]
\end{equation}
The plot of effective potential is shown in fig. (\ref{VeffWCF}).
To describe the shape of the orbit, we need to find the relationship between the azimuthal angle $\phi$ and the radial coordinate r to see how $\phi$ changes with respect to r. We get,
\begin{equation}
    \frac{d\phi}{dr} = \frac{l}{r^2} \hspace{0.1cm} \sqrt{\frac{1}{2(Veff-E)}}
\end{equation}
To simplify the above equation, let $u = \frac{1}{r}$. Then the orbit equation is the second order differential equation by differentiating the above equation with respect to $\phi$. The orbit equation obtained is,
\begin{equation}
    \frac{d^2 u}{d\phi^2} + u (1 + Mu)^{-2} - Mu^2 (1 + Mu)^{-3} - \frac{M}{l^2} (1 + Mu)^{-3} = 0
\end{equation}
The particle orbit for the naked singularity space-time in eq.(\ref{naked}) is given in fig. (\ref{WCFPlot}).
\subsection{For Regular Naked Singularity Space-time}
For the regularized space-time in eq.(\ref{regular}), using the conserved $\gamma$ and l, we get,
\begin{equation}
\begin{split}
    u^t = \frac{\gamma}{\Big(1 + \frac{L^2}{r^2}\Big) \Big(1 + \frac{M}{r}\Big)^{-2}} \\
   and \hspace{0.2cm} u^\phi = \frac{l}{r^2 \Big(1 + \frac{L^2}{r^2}\Big)}
\end{split}
\end{equation}
Also here, we have considered equatorial region, so, $\theta = \frac{\pi}{2}$.\\
The effective potential is,
\begin{equation}
    \frac{1}{2}\Big[\frac{l^2}{r^2} \Big(1 + \frac{M}{r}\Big)^{-2} + \Big(1 + \frac{L^2}{r^2}\Big) \Big(1 + \frac{M}{r}\Big)^{-2} - 1 \Big]
\end{equation}
The plot for effective potential for the regularized space-time is given in fig. (\ref{VeffCF}).
For this case, the orbit equation obtained is,
\begin{equation}
\begin{split}
    \frac{d^2u}{d\phi^2} + u(1+Mu)^{-2} - Mu^2(1+Mu)^{-3} +\\ \frac{L^2}{l^2} u (1+Mu)^{-2} - \frac{M}{l^2} (1+L^2u^2) (1+Mu)^{-3} = 0
    \end{split}
\end{equation}
The orbit plot for the above regularized space-time is given in fig. (\ref{CFPlot}).

\section{Precession of Bound Orbits in Singular and Regular Naked Singularity Space-times}
The orbit equations for the given space-times provide us information about the shapes of the particle orbits. Now, to understand the precession of the particle orbits, we can find approximation solutions for these orbit equations. One of the approximations we will use here is the low eccentricity approximation, considering the eccentricity of the orbit to be very small so that the higher order eccentricity terms are neglected. \\
The low eccentricity approximation solution for the orbit is given by,
\begin{equation} \label{u}
    u = \frac{1}{p} [1 + e \cos(m\phi)]
\end{equation}
where p and m are positive real numbers.\\
The first order derivative of u is,
\begin{equation}
    u' = -\frac{em}{p} \sin(m\phi)
\end{equation}
and the second order derivative of u is,
\begin{equation} \label{u''}
    u'' = -\frac{em^2}{p} \cos(m\phi)
\end{equation}
Using eq (\ref{u}), (\ref{u''}) and the orbit equation, we can determine p and m.

\subsection{Naked Singularity Space-time}
\begin{figure*}
    \subfigure[$m>1$]{ \includegraphics[scale=0.5]{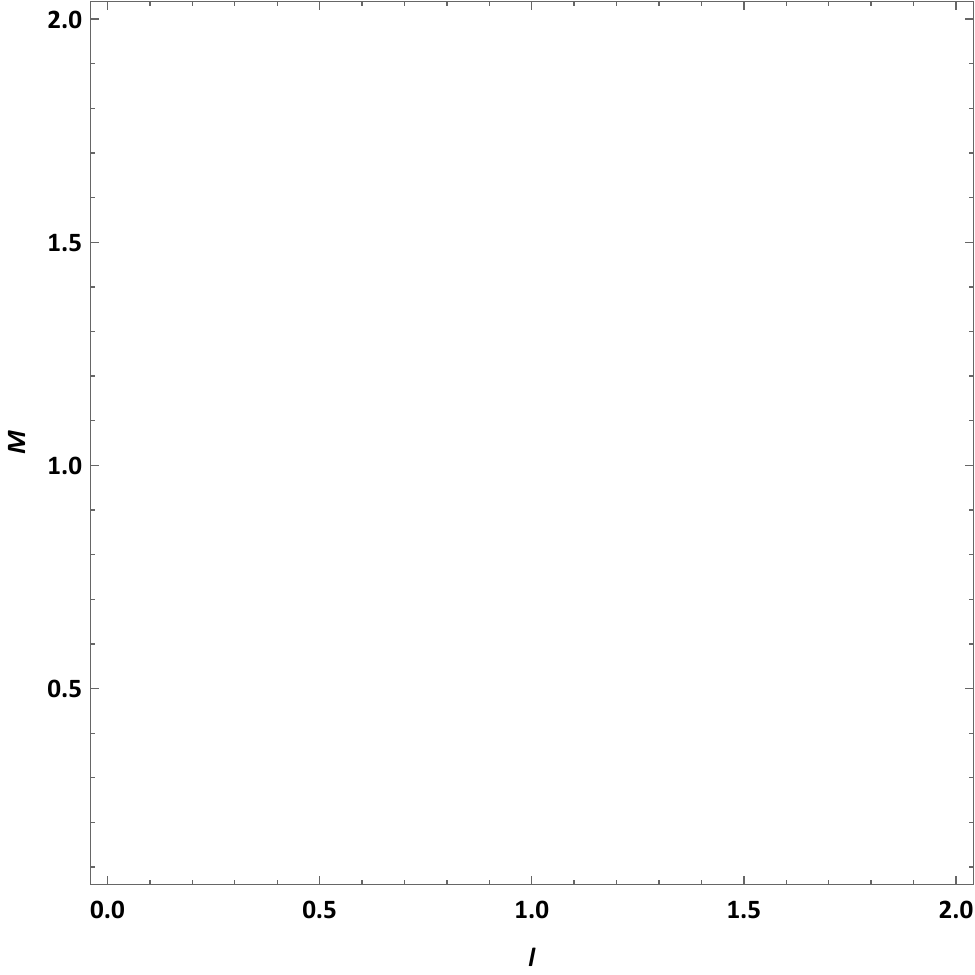}\label{negwcf}}
     \hspace{0.3cm}
     \subfigure[$m<1$]{ \includegraphics[scale=0.5]{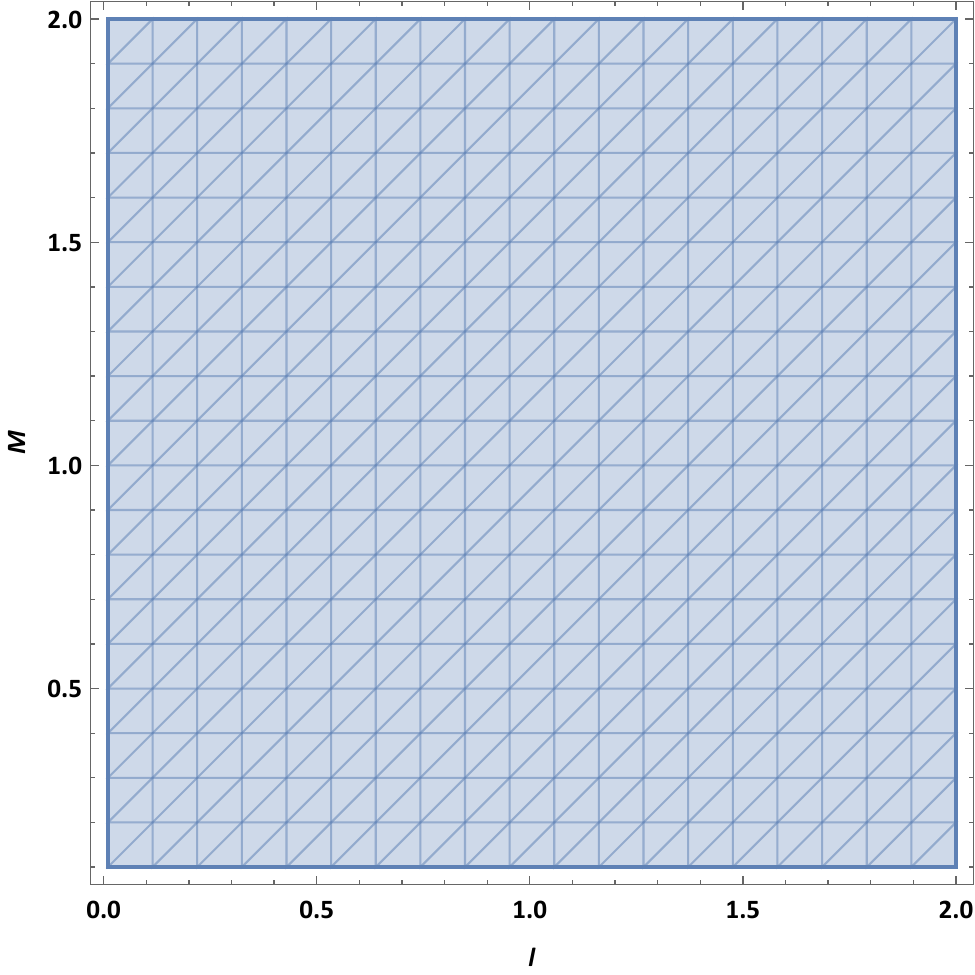}\label{poswcf}}
     \caption{Negative and positive precession for naked singularity space-time.}
\end{figure*}

\begin{figure}[h]
    \includegraphics[scale=0.5]{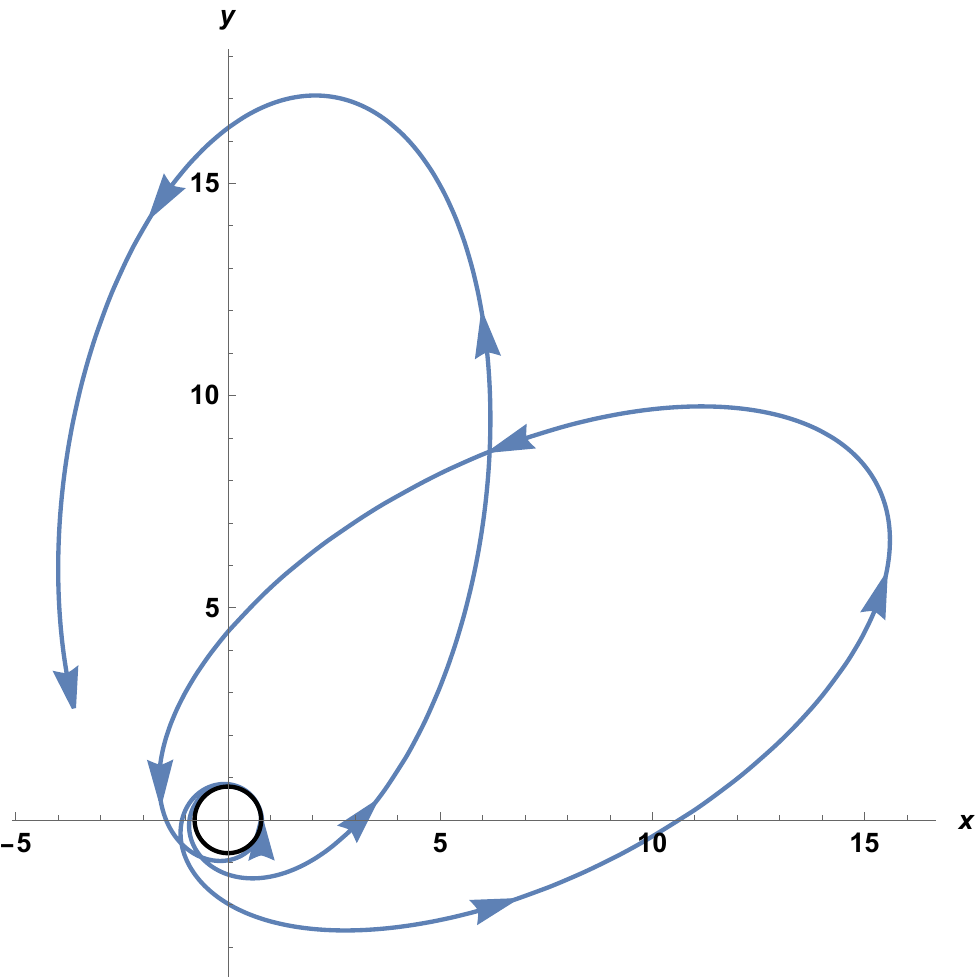}
    \caption{Positive precession in the orbit of naked singularity space-time for $M=1$, $l=1.5$. The black circle shows the minimum approach of the orbit.}
    \label{posorbitwcf}
\end{figure}

Using eq (\ref{naked}), (\ref{u}), (\ref{u''}) and neglecting the higher order terms, we obtain the following expressions for p and m,
\begin{equation}
    p = \frac{l^2}{M} \hspace{0.3cm} and
\end{equation}
\begin{equation}
    m = -i \sqrt{\frac{l^2}{M}} \hspace{0.1cm} \sqrt{-\frac{l^6}{M^3 \big(\frac{l^2}{M} + M\big)^4} - \frac{l^4}{M \big(\frac{l^2}{M} + M\big)^4}}
\end{equation}
For $m<1$ and $m>1$, the range of l and M is given in fig. (\ref{negwcf}) and (\ref{poswcf}) respectively.
Here, for any value of l or M, we can observe that orbits are only possible for $0<m<1$. This shows that starting from the closest approach to the center, after completing a rotation of 2 $\pi$, the particle does not reach its previous position. To reach its previous position it needs to gain some extra angle which depends upon the value of m. We can see that the orbit precesses in the same direction as the rotation of the particle. This shows that this naked singularity space-time possesses positive precession in its orbits. The orbit is shown in fig.(\ref{posorbitwcf}).

\subsection{Regularized Naked Singularity Space-time}
\begin{figure*}
    \centering
    \subfigure[$m>1, L=0.3$]{\includegraphics[scale=0.38]{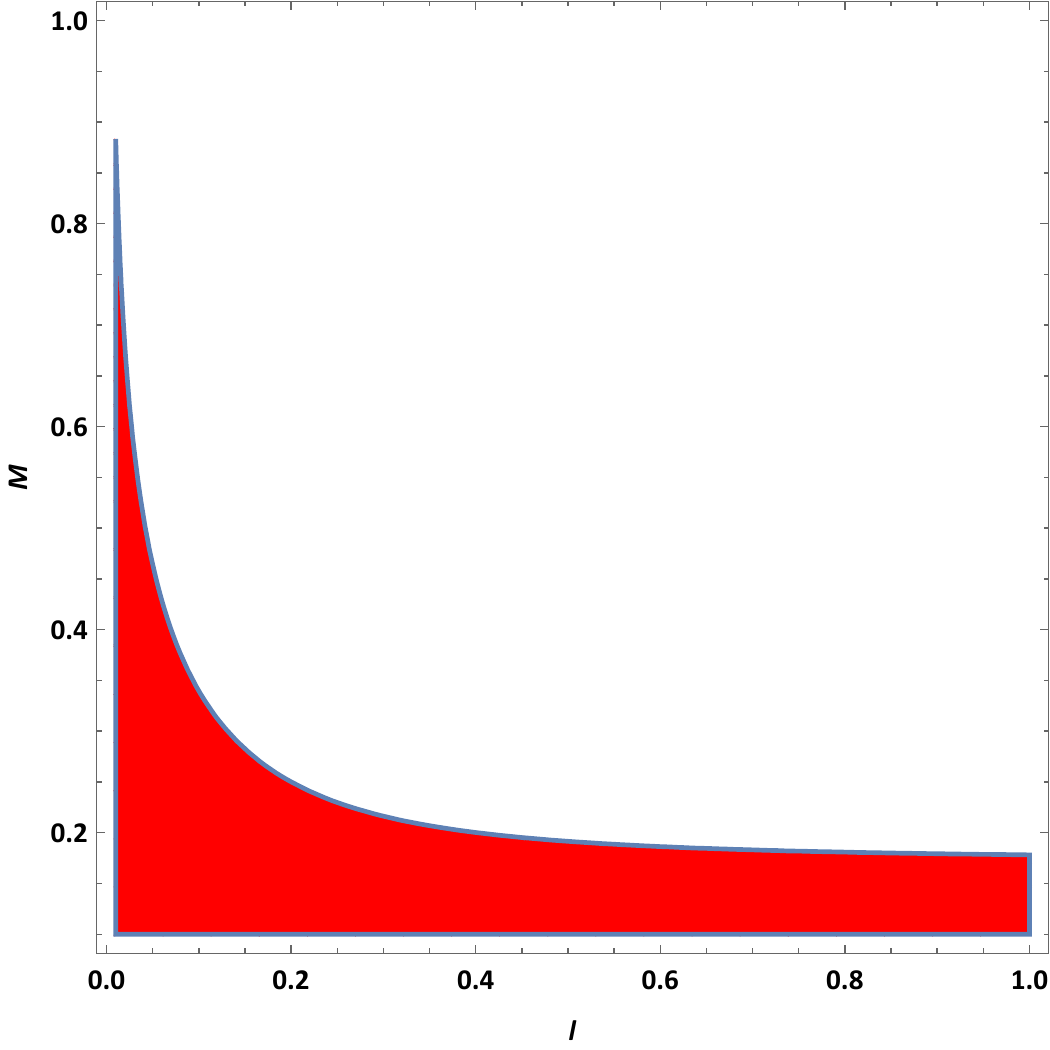}\label{a}}
    \hspace{0.2cm}
    \subfigure[$m<1, L=0.3$]{\includegraphics[scale=0.38]{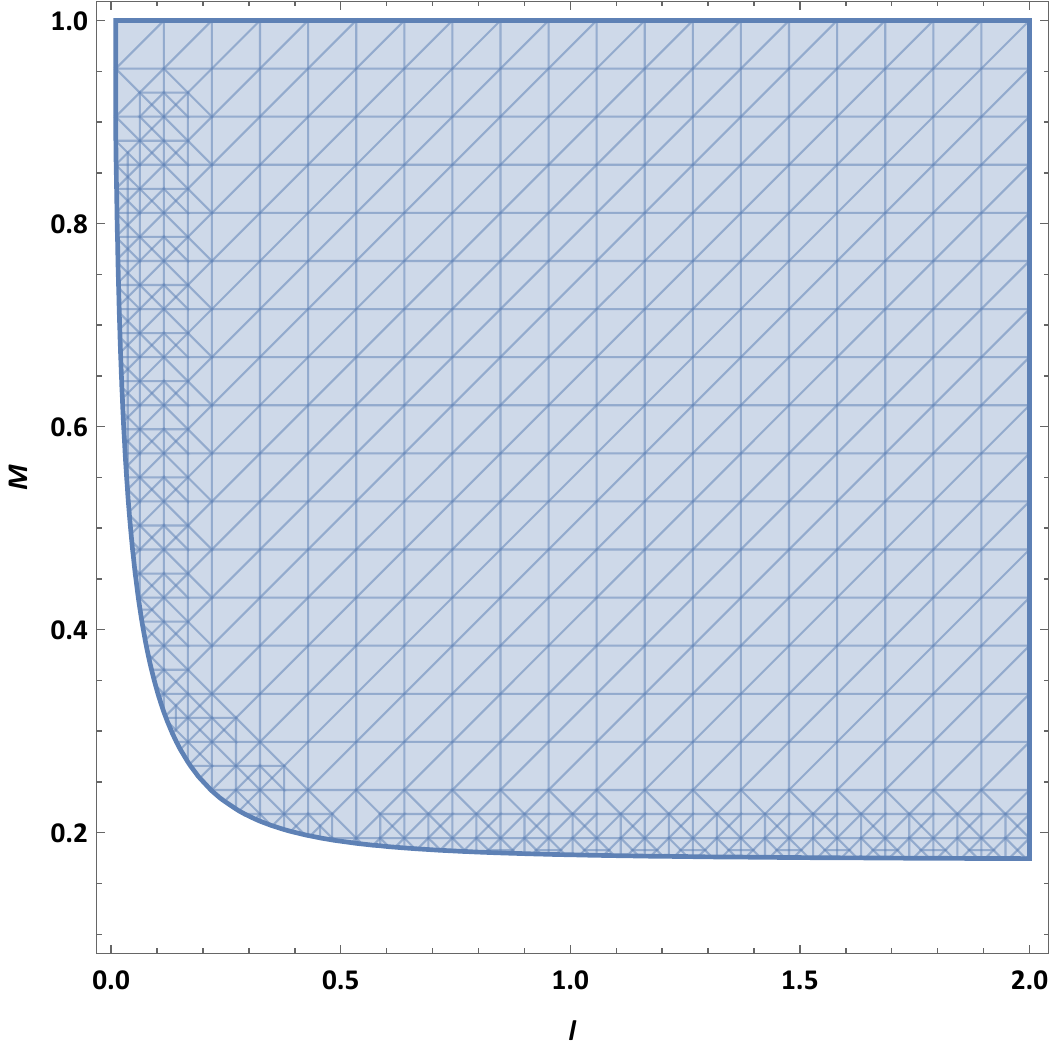}\label{b}}
    \hspace{0.2cm}
    \subfigure[$m>1, L=0.5$]{\includegraphics[scale=0.38]{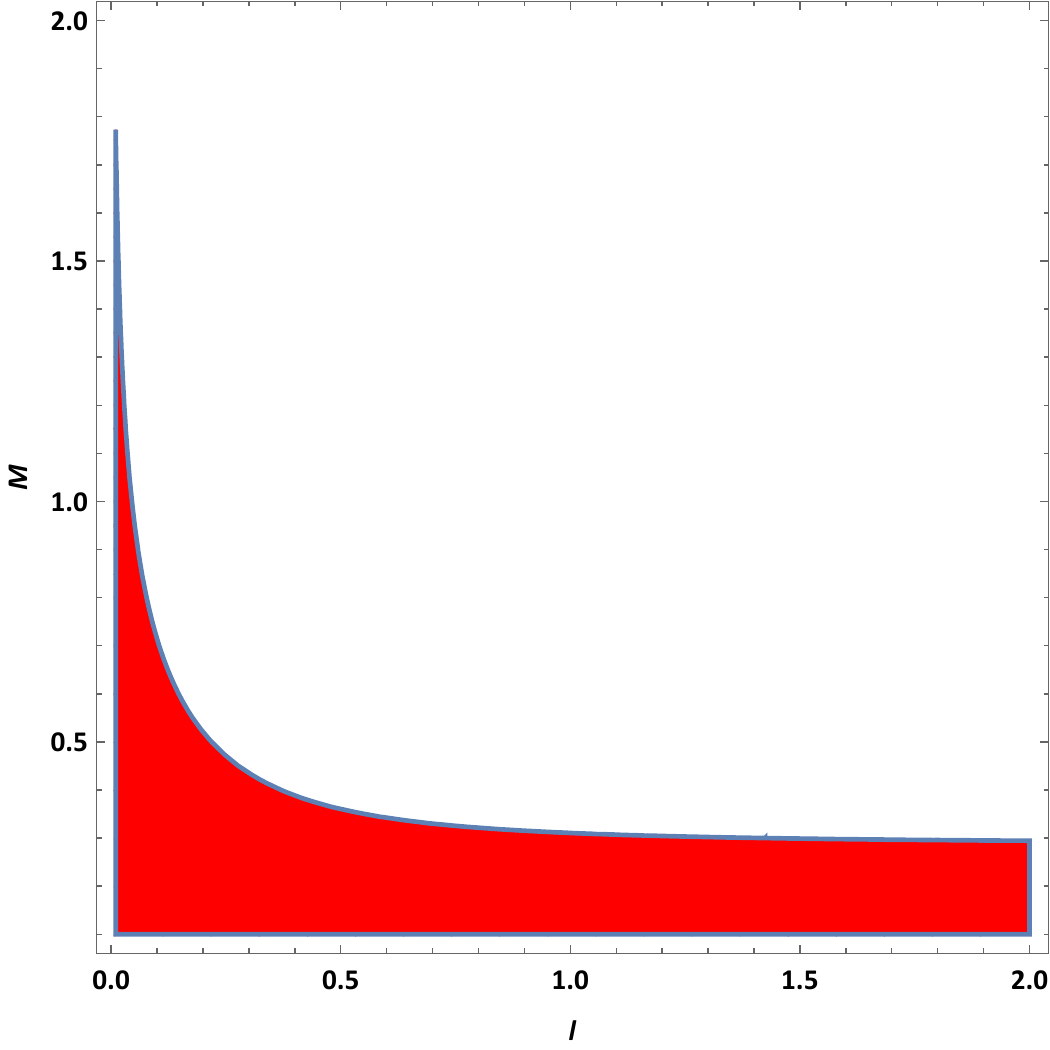}\label{c}}
    \hspace{0.2cm}
    \subfigure[$m<1, L=0.5$]{\includegraphics[scale=0.38]{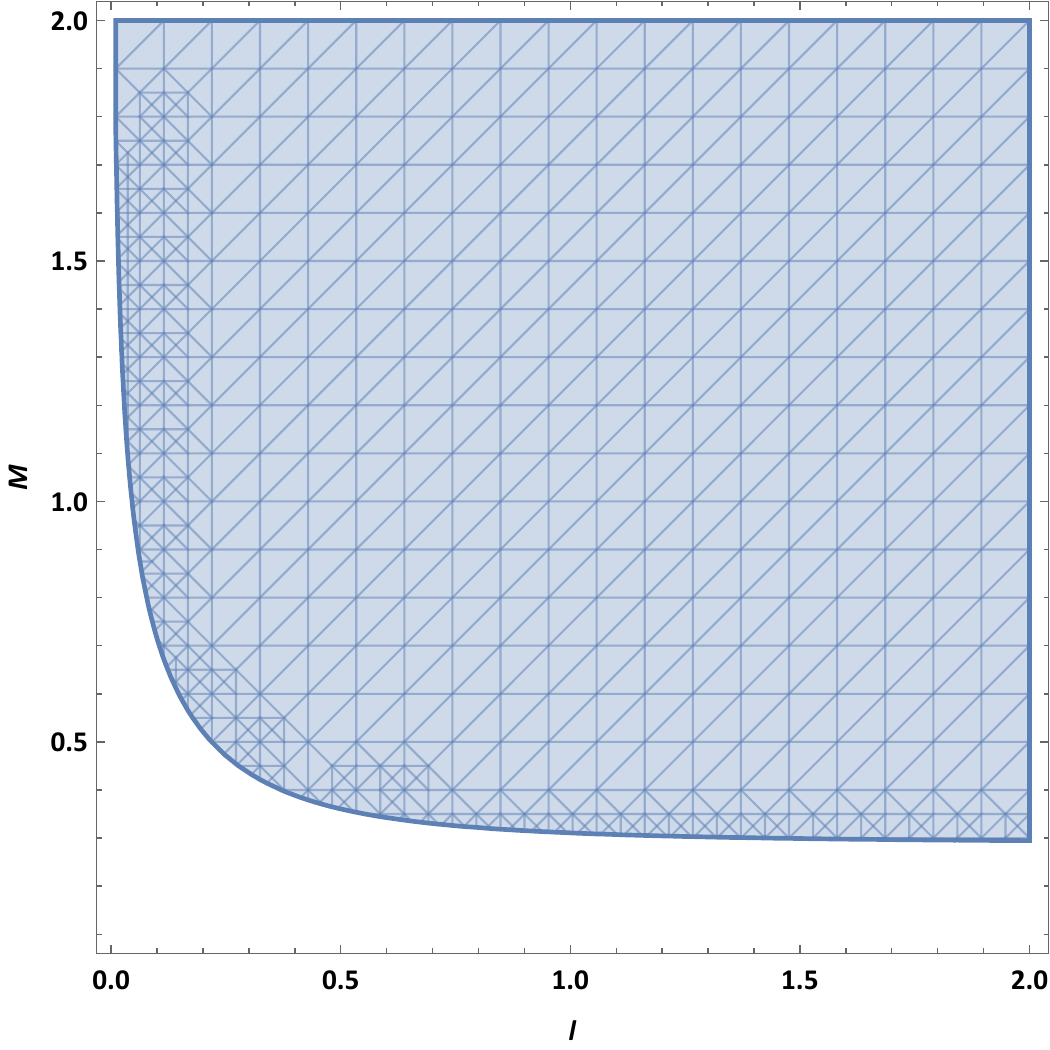}\label{d}}  \hspace{0.2cm}
    \subfigure[$m>1, L=1$]{\includegraphics[scale=0.38]{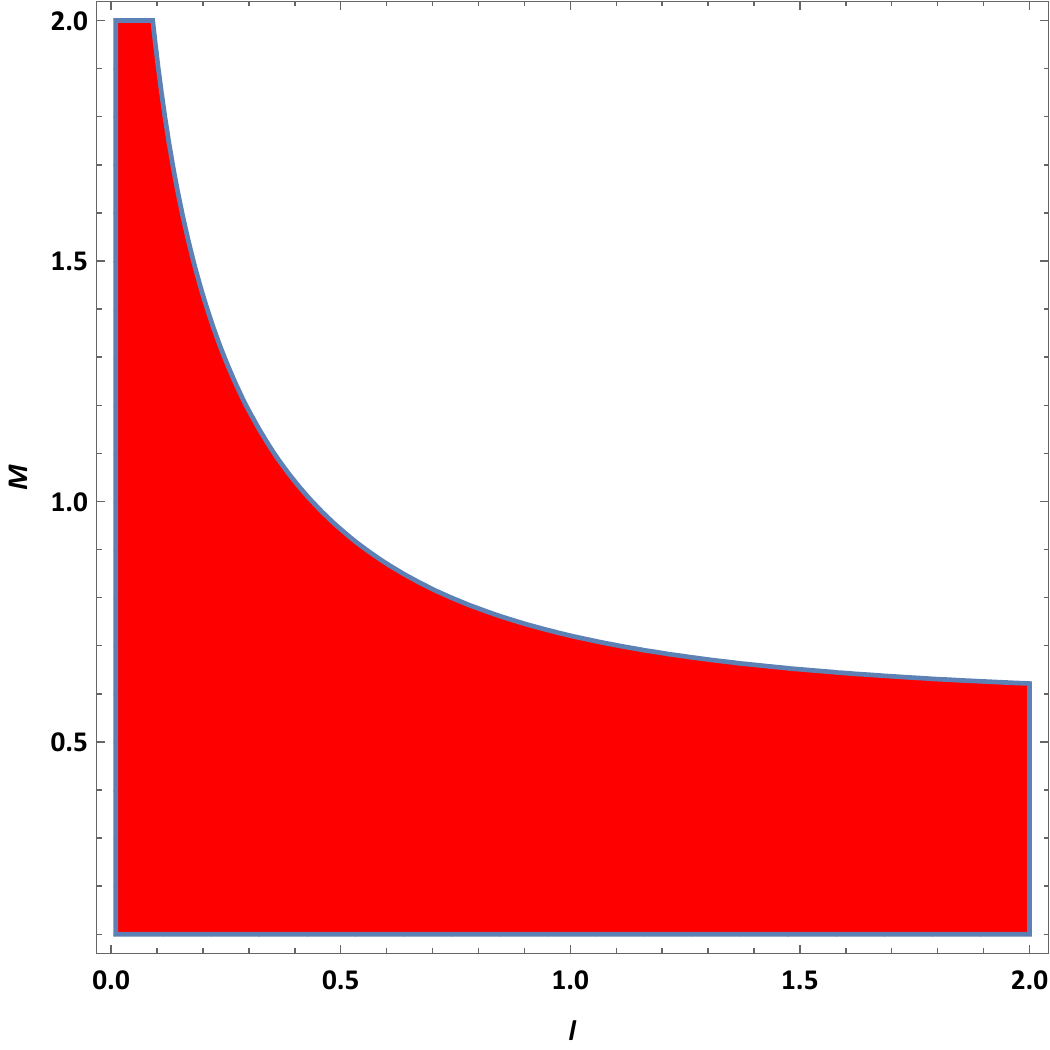}\label{e}}
    \hspace{0.2cm}
    \subfigure[$m<1, L=1$]{\includegraphics[scale=0.38]{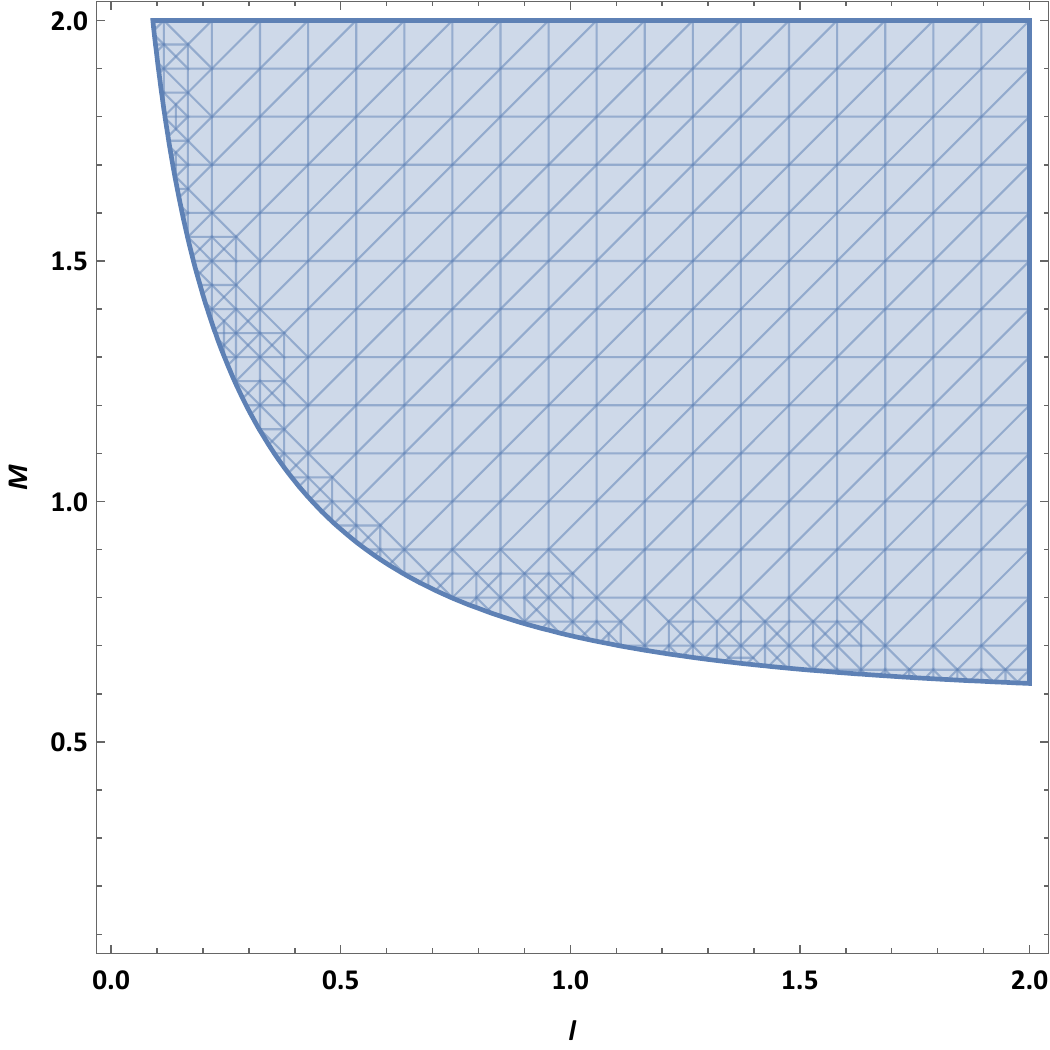}\label{f}}
    \caption{The above figures show how the range of negative and positive precession changes for different L}
\end{figure*}

Using eq (\ref{regular}), (\ref{u}), (\ref{u''}), the expressions for p and m are,
\begin{equation}
    p = \frac{l^2 + L^2}{M} \hspace{0.3cm} and
\end{equation}
\begin{multline}
    m = -i \sqrt{\frac{l^2 + L^2}{M}} \hspace{0.1cm} \Bigg(-\frac{(l^2+L^2)^3}{M^3 \Big(\frac{l^2 + L^2}{M} + M\Big)^4} - \\ \frac{L^2 (l^2+L^2)^3}{l^2 M^3 \Big(\frac{l^2 + L^2}{M} + M\Big)^4} + \frac{2 (l^2+L^2)^2}{M \Big(\frac{l^2 + L^2}{M} + M\Big)^4} \\ + \frac{2 L^2 (l^2+L^2)^2}{l^2 M \Big(\frac{l^2 + L^2}{M} + M\Big)^4} - \frac{3 (l^2+L^2)^3}{l^2 M \Big(\frac{l^2 + L^2}{M} + M\Big)^4}\Bigg)^{\frac{1}{2}}
\end{multline}
Here, we can observe that both $m>1$ and $0<m<1$ are possible. Let us see how the range of $m>1$ and $m<1$ changes in terms of l and M for different values of L:
\begin{enumerate}
    \item $\underline{L=0.3}$\\
    For $m>1$ and $m<1$, the range of l and M is given in fig.(\ref{a}) and (\ref{b}).
    \item $\underline{L=0.5}$\\
    The range of l and M is provided in fig.(\ref{c}) and (\ref{d}).
    \item $\underline{L=1}$\\
    The range of l and M is provided in fig.(\ref{e}) and (\ref{f}).
\end{enumerate}

\begin{figure*} 
    \centering
    \subfigure[Negative precession for $m>1$, $L=0.3$, $l=0.16$, $M=0.2$]{\includegraphics[scale=0.4]{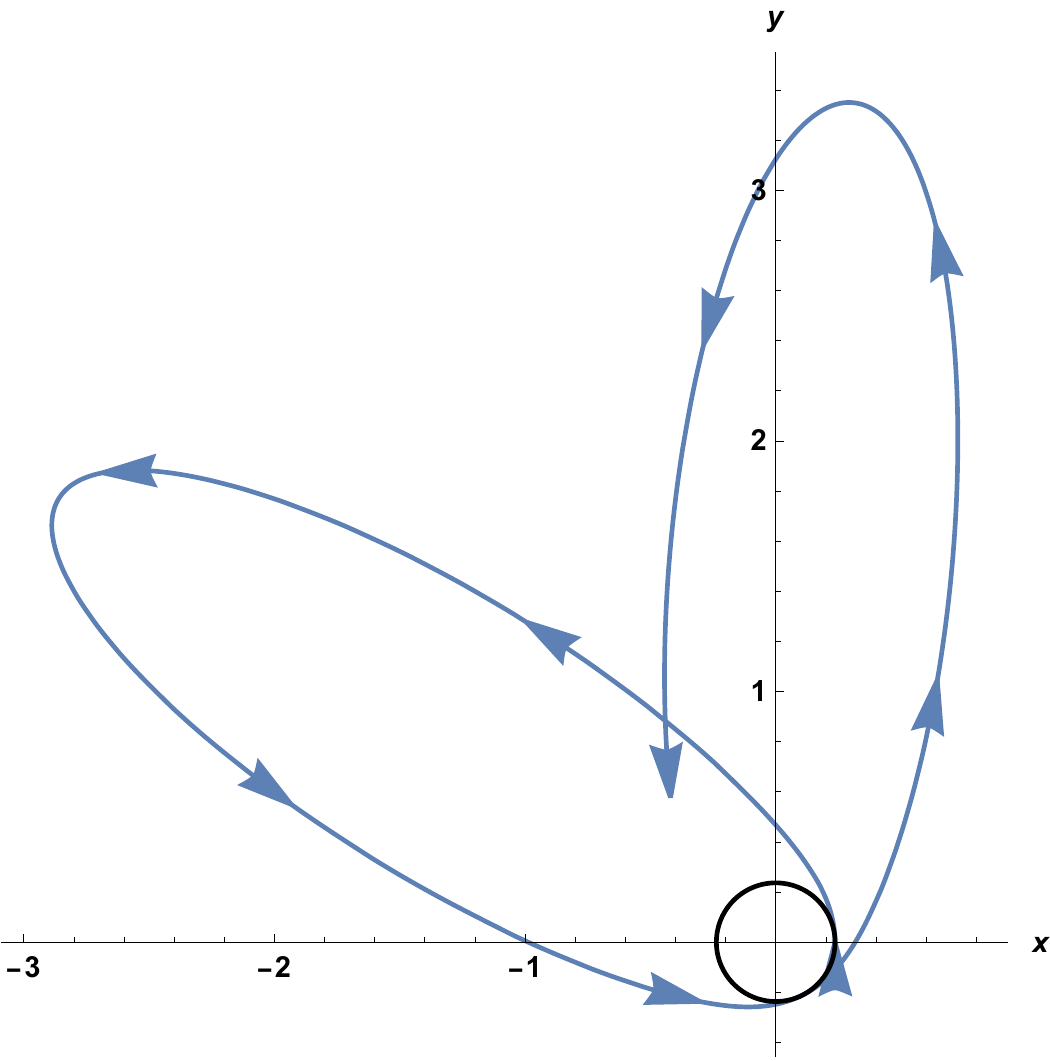}\label{A}} \hspace{0.5cm}
    \subfigure[Positive precession for $m<1$, $L=0.3$, $l=1$, $M=0.2$]{\includegraphics[scale=0.4]{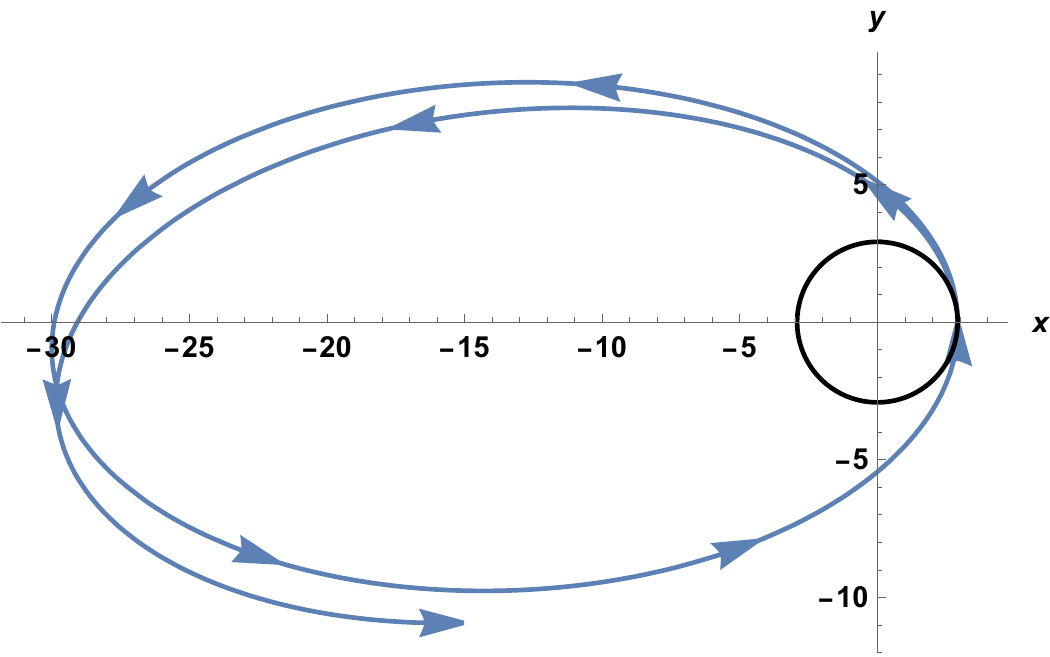}\label{B}} \hspace{0.5cm}
    \subfigure[Negative precession for $m>1$, $L=0.5$, $l=0.08$, $M=0.5$]{\includegraphics[scale=0.4]{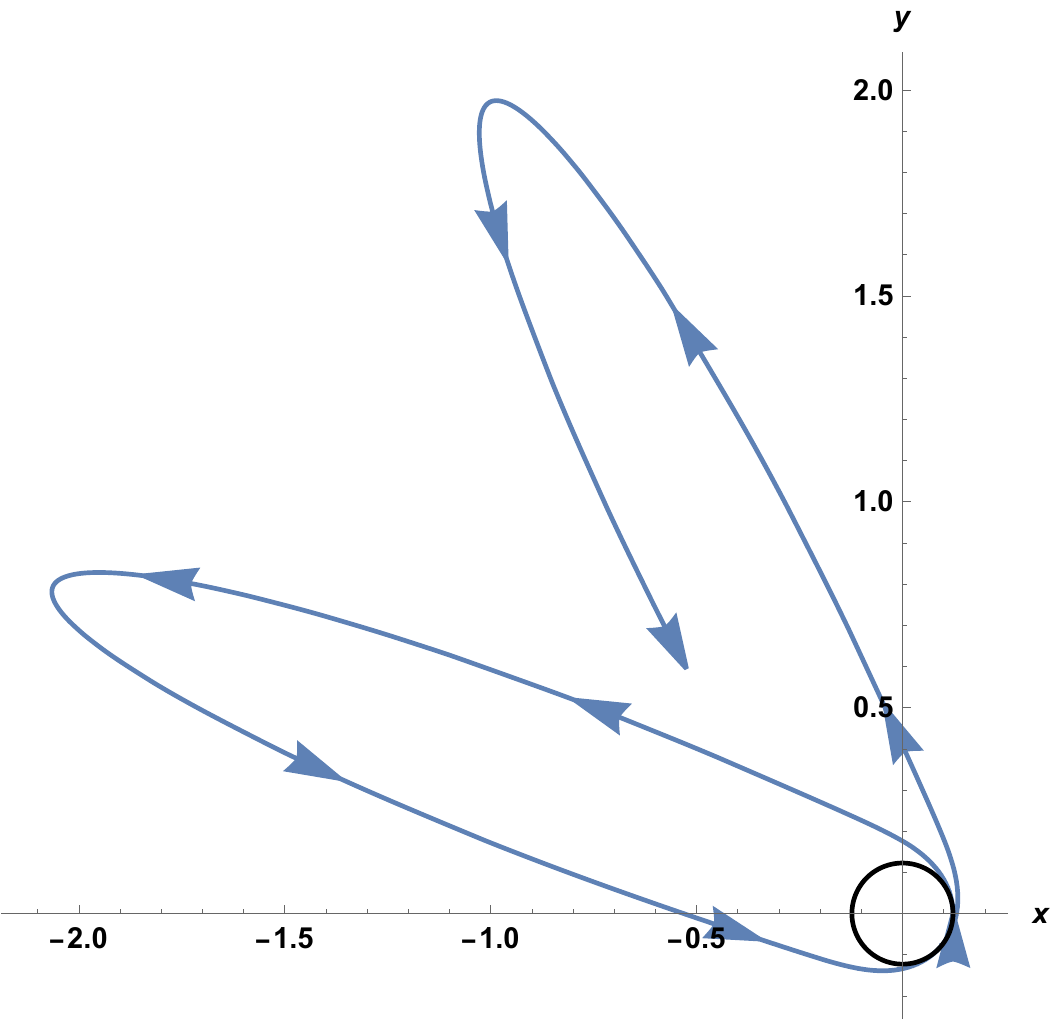}\label{C}} \hspace{0.5cm}
    \subfigure[Positive precession for $m<1$, $L=0.5$, $l=1.6$, $M=0.5$]{\includegraphics[scale=0.4]{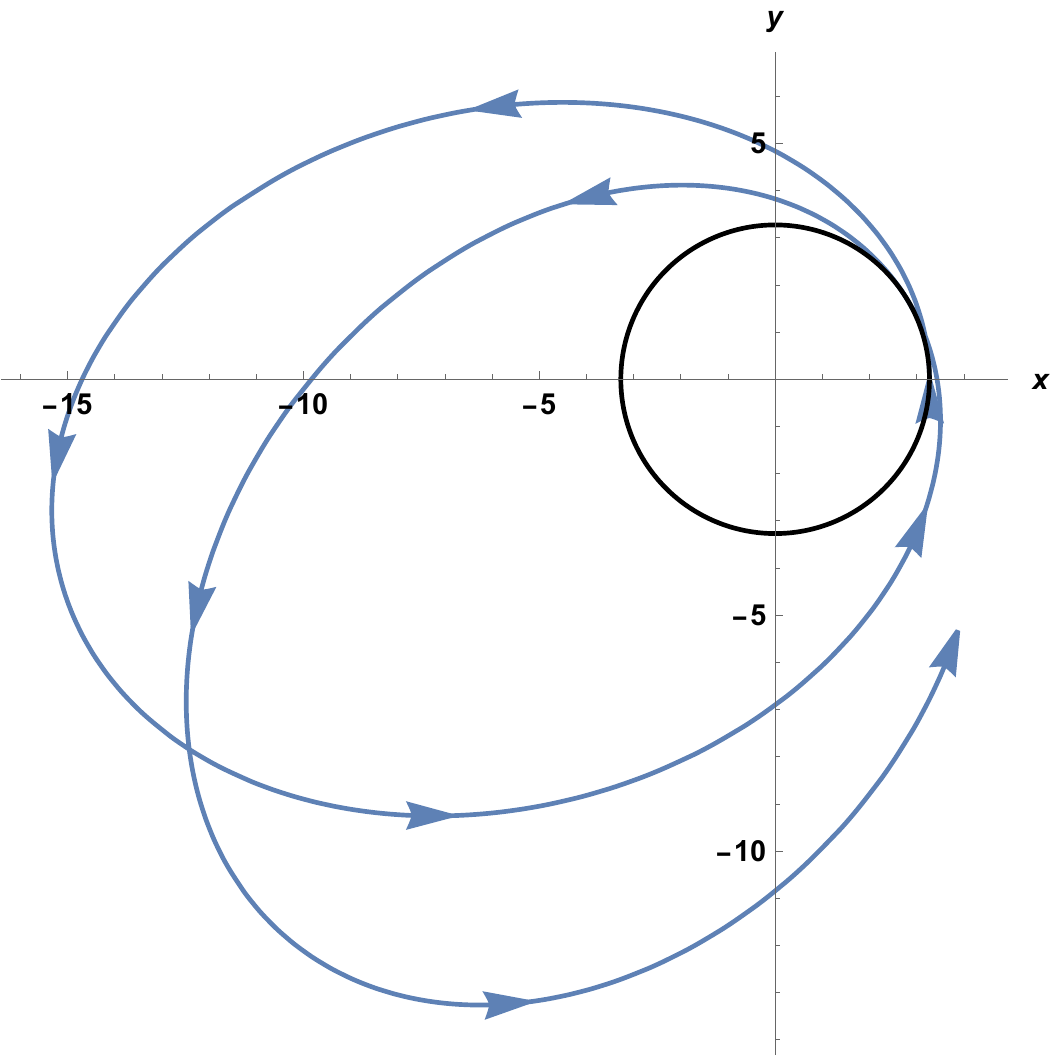}\label{D}} \hspace{0.5cm}
    \subfigure[Negative precession for $m>1$, $L=1$, $l=0.3$, $M=1$]{\includegraphics[scale=0.4]{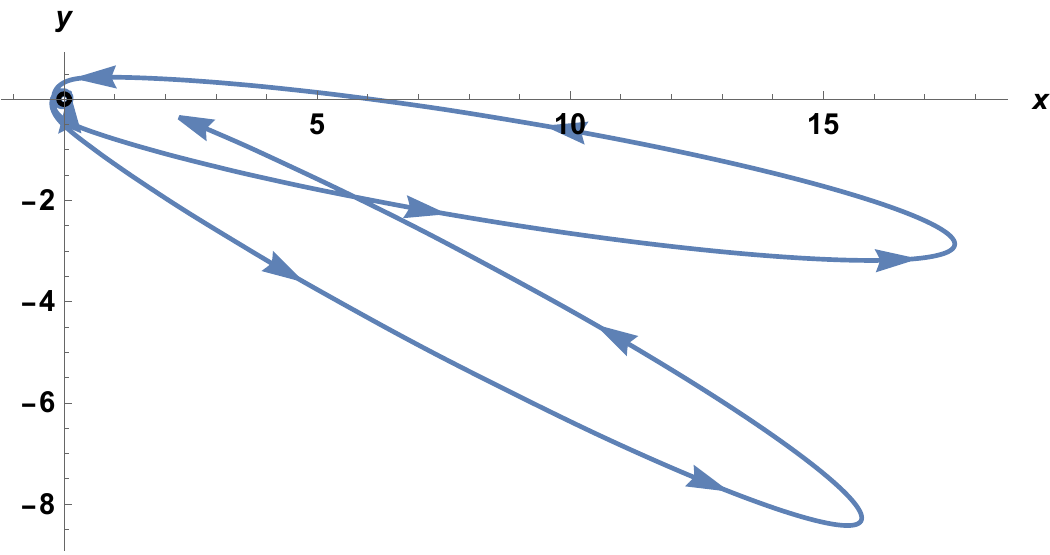}\label{E}} \hspace{0.5cm}
    \subfigure[Positive precession for $m<1$, $L=1$, $l=1.7$, $M=1$]{\includegraphics[scale=0.4]{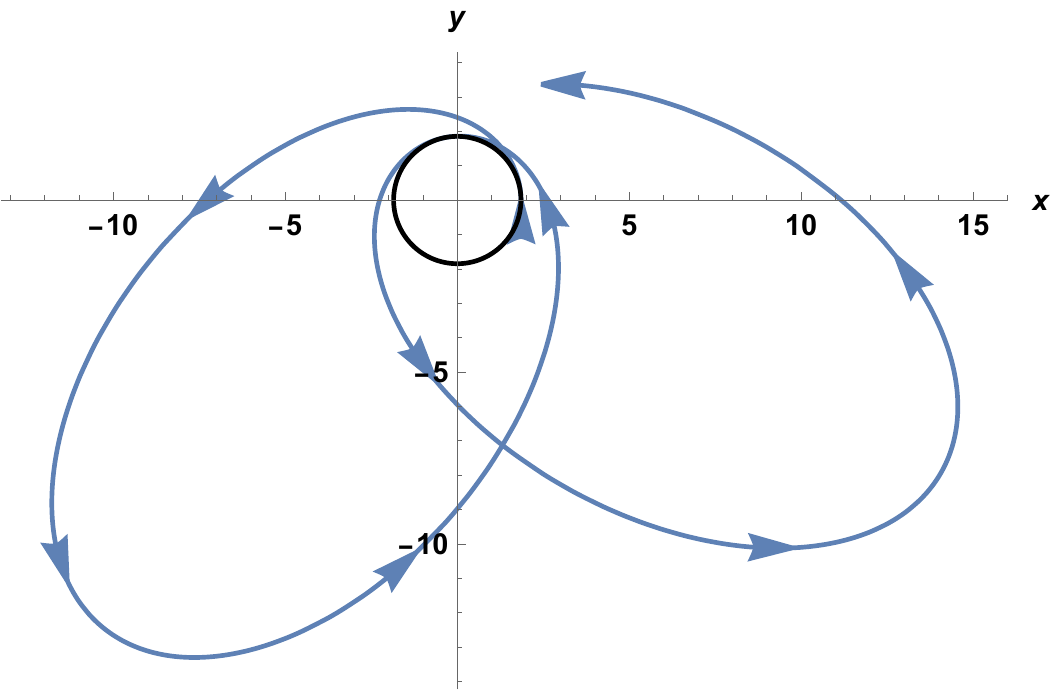}\label{F}}
\caption{Shape of bound orbit plots for different L values in regular naked singularity space-time. Here, all the black outlines show the closest approach of the orbits.}
\label{regorbit}
\end{figure*}

The orbits for different L values are shown in fig.(\ref{regorbit}). We can observe that in the case of the regular naked singularity space-time, orbits are possible for both $m<1$ and $m>1$. \\
In fig.(\ref{A}), (\ref{C}) and (\ref{E}), we can see that the orbit precesses in the opposite direction as the direction of the rotation of the particle. Thus, this gives the negative precession of the particle orbits.\\
Whereas in fig.(\ref{B}), (\ref{D}) and (\ref{F}), it is seen that the direction of orbit precession is the same as that of the rotation of the particle. Thus, this gives the positive precession of the particle. Therefore, we can observe that the regular naked singularity space-time can have both negative and positive precession depending upon the values of L, l and M.

\section{Discussion and Conclusions}
In this study, we regularized a naked singularity space-time using conformal transformations. Then we examined the time-like trajectories of particles and studied the differences between the orbital dynamics of the singular and regular naked singularity space-times which are conformally connected. The study of these particle orbits brought some interesting differences between the singular and regular space-times which are summarized below:
\begin{enumerate}
    \item For the naked singularity space-time given in eq.(\ref{naked}), only positive precession of the particle orbits is possible as given in fig.(\ref{posorbitwcf}). But for the regular naked singularity space-time in eq.(\ref{regular}), there is a possibility of both negative and positive precession in the orbits which can be determined by the values of l, M and L (as shown in fig.(\ref{regorbit})).
    \item We can observe that smaller the value of L, smaller becomes the range of $m>1$ and larger becomes the range of $m<1$, i.e., the range of negative precession of orbits becomes smaller while that for positive precession becomes larger as we approach nearer to the center. But, the effect of negative precession will still be prominent nearer to the center and that of the positive precession will be prominent away from the center.
    \item In the case of negative precession $m>1$, for $l<<L$, the shape of the particle orbits become more elongated. This can be observed in fig. (\ref{C}) and (\ref{E}).
    \item Both the positive precession orbits in figures (\ref{posorbitwcf}) and (\ref{F}) have $M=1$ and $l=1.5$, but their orbit orientations are different due to the inclusion of the conformal factor given in eq.(\ref{confactor}) in fig.(\ref{F}) which makes the positive precession orbits distinguishable from each other.
    \item We obtained the results of the orbits in regular naked singularity due to the specific choice of the conformal factor, i.e.,
    \begin{equation}
        S(r) = \Big(1+\frac{L^2}{r^2}\Big)
    \end{equation}
    But, as there can be more than one conformal factor to regularize a space-time \cite{bambi}, \cite{pn}, so the nature of the orbits might be different for a different choice of conformal factor S(r). But, even those orbits of a regularized space-time formed by a different choice of S(r) would differ from the orbits of the singular space-time due to the influence of the conformal factor S(r).
\end{enumerate}
In this paper, we have shown that the orbital dynamics of a singular and regular naked singularity space-time can be significantly different from each other. As we know that many missions are targeted to observe the stellar motions of the S-stars around the galactic centers of our own Milky Way galaxy and other galaxies, these observations can give us significant information about the orbital dynamics of the stars around the central compact objects and further be helpful in leading us to the nature of those compact objects.
\begin{center}
    Acknowledgement
\end{center}
I would like to thank my mentor Dr. Dipanjan Dey for guiding me through this work by engaging in insightful discussions and suggestions.

\end{document}